\documentclass[notitlepage,aps,prl,showpacs,singlecolumn,amsmath,amssymb,superscriptaddress]{revtex4-1}

\usepackage{graphicx}
\usepackage{dcolumn}
\usepackage{bm}
\usepackage{units}
\usepackage{upgreek}
\usepackage{soul}
\usepackage{ucs}
\usepackage{color}
\usepackage{cancel}
\usepackage{braket}
\usepackage{hyperref}
\usepackage[toc,page]{appendix}
\usepackage[all]{hypcap}
\usepackage{textcomp}
\usepackage{color}
\usepackage{soul}

\preprint{APS/123-QED}

\begin{document}

\title{Measuring finite-range phase coherence in an optical lattice using \\ Talbot interferometry}
\author{Bodhaditya Santra}
\affiliation{Department of Physics and Research Center OPTIMAS, Technische Universit\"at Kaiserslautern, 67663 Kaiserslautern, Germany}
\author{Christian Baals}
\affiliation{Department of Physics and Research Center OPTIMAS, Technische Universit\"at Kaiserslautern, 67663 Kaiserslautern, Germany}
\affiliation{Graduate School Materials Science in Mainz, Staudinger Weg 9, 55128 Mainz, Germany}
\author{Ralf Labouvie}
\affiliation{Department of Physics and Research Center OPTIMAS, Technische Universit\"at Kaiserslautern, 67663 Kaiserslautern, Germany}
\affiliation{Graduate School Materials Science in Mainz, Staudinger Weg 9, 55128 Mainz, Germany}
\author{Aranya B. Bhattacherjee}
\affiliation{School of Physical Sciences, Jawaharlal Nehru University, New Delhi-110067, India}
\author{Axel Pelster}
\affiliation{Department of Physics and Research Center OPTIMAS, Technische Universit\"at Kaiserslautern, 67663 Kaiserslautern, Germany}
\author{Herwig Ott \footnote{Correspondence and requests for materials should be addressed to H.O. (ott@physik.uni-kl.de)}}
\affiliation{Department of Physics and Research Center OPTIMAS, Technische Universit\"at Kaiserslautern, 67663 Kaiserslautern, Germany}
\email{ott@physik.uni-kl.de}

\begin{abstract}

One  of  the  important  goals  of  present  research is to control and manipulate coherence in a broad variety of systems, such as semiconductor spintronics, biological photosynthetic systems, superconducting qubits and complex atomic networks. Over the past decades interferometry of atoms and molecules has proven to be a powerful tool to explore coherence. Here we demonstrate a near-field interferometer based on the Talbot effect, which allows to measure finite-range phase coherence of ultracold atoms in an optical lattice. We apply this interferometer to study the build-up of phase coherence after a quantum quench of a Bose-Einstein condensate residing in a one-dimensional optical lattice. Our technique of measuring finite-range phase coherence is generic, easy to adopt, and can be applied in practically all lattice experiments without further modifications.

\end{abstract}



\maketitle

{\large {\bf Introduction}} 

First- and second-order correlations are among the most important observables in ultracold quantum gas experiments. 
First-order correlations reflect the phase coherence between the atoms and are visible in the contrast of interference experiments \cite{Andrews1997,Greiner2001} in time of flight expansion.
For long expansion times, all atoms interfere with each other and the contrast of the interference pattern measures the global phase coherence \cite{Gerbier2005,Braun2015}. 
Second-order correlations can be accessed via Bragg scattering \cite{Ozeri2005} or the measurement of the density fluctuations. 
The latter has been demonstrated for time of flight absorption images \cite{Foelling2005}, the detection of metastable atoms \cite{Vassen2012,Jeltes2007} and single atom sensitive {\it in situ} detection methods \cite{Ott2016,Guarrera_2012,Cheneau2012}. 

A very peculiar method to measure first-order correlations has been developed for one-dimensional quantum gases. 
There, the interference with a twin system \cite{Hofferberth2007} is used to investigate the local phase evolution of the gas. 
Analyzing the full distribution function of the interference contrast \cite{Hofferberth2008}, it is possible to study non-equilibrium many-body dynamics such as the appearance of prethermalization in isolated, one-dimensional quantum systems \cite{Gring2012} or the properties of generalized Gibbs ensembles \cite{Langen2015}. 
A similar experimental approach in two dimensions has also recently been developed \cite{Chomaz2015}. 
Related physical questions in three-dimensional optical lattices have been addressed via the measurement of the global coherence of the matter wave interference pattern and the emergence of phase coherence after a quantum quench \cite{Braun2015}. The advent of quantum gas microscopy \cite{Bakr2010,Sherson2010,Ott2016} has established a superb tool to measure the density distribution and second-order correlation functions in optical lattices with single-site resolution. 
However, a complementary protocol for single-site resolved measurements of first-order correlations is so far missing.

Here, we describe a Talbot interferometer, which is capable to probe the phase coherence of ultracold atoms in an optical lattice for well defined distances.
We experimentally demonstrate this interferometer by measuring the build up of phase coherence after a quantum quench of a Bose-Einstein condensate (BEC) in a one-dimensional optical lattice \cite{Cataliotti2001}. 
The interferometric protocol is applicable in practically any optical lattice experiment without further modification. 
For high resolution {\it in situ} imaging techniques, the measurement principle can be combined with spatially resolved readout, thus paving the way to locally probe finite-range phase coherence in many-body quantum systems.\\

\begin{figure}[t]
\begin{center}
\includegraphics[width=8.5cm,angle=0]{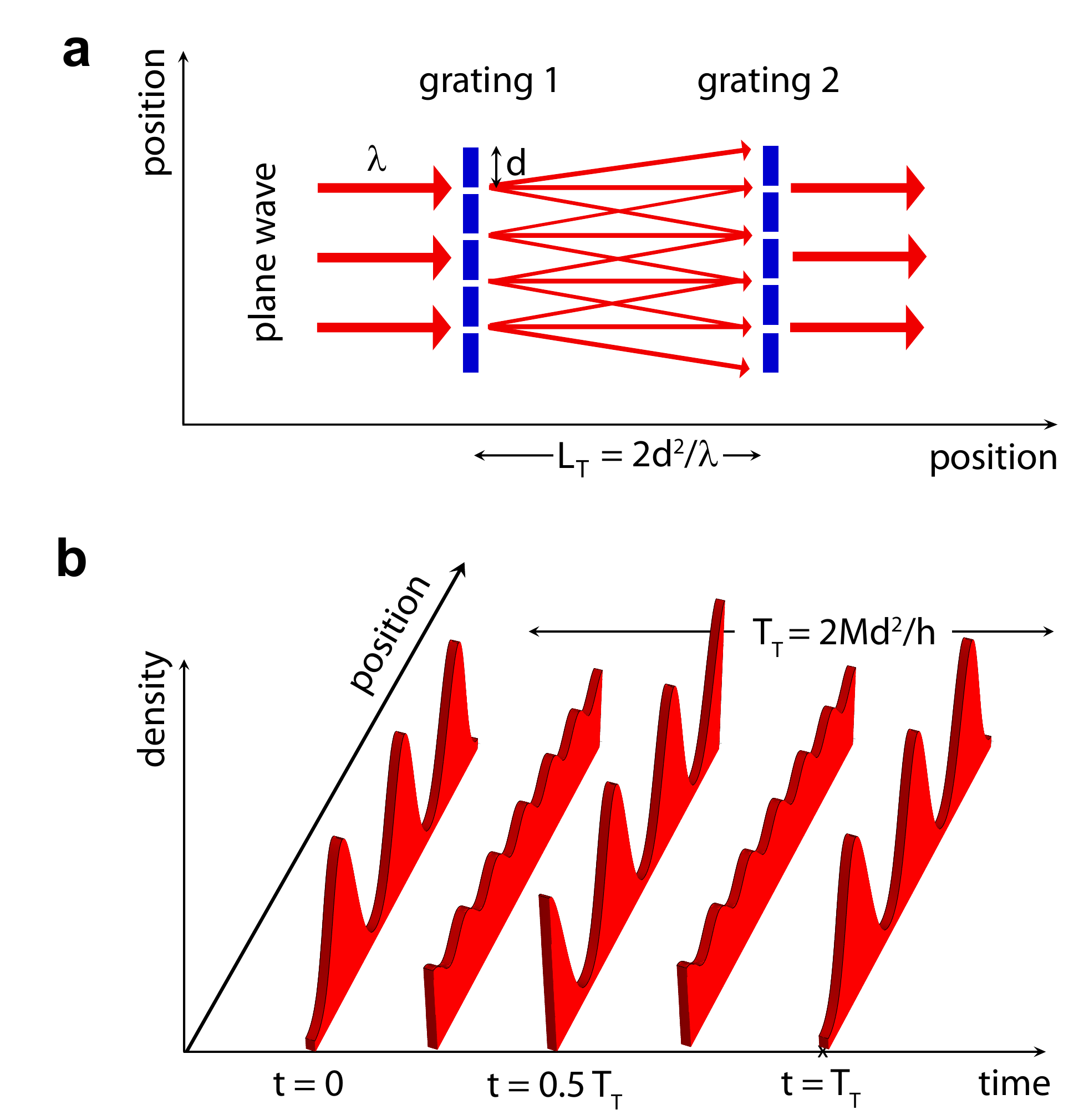}
\end{center}
\caption{{\bf Spatial and temporal Talbot effect} {\bf a} A plane wave with wavelength $\lambda$ passes a grating with lattice constant $d$. At integer multiples of the Talbot distance $L_{\rm T}=2d^2/\lambda$, the interference pattern shows revivals. These revivals are probed by the transmission of a second identical grating. {\bf b} A coherent matter wave is trapped in a periodic potential and starts to interfere after switching off the potential. After integer multiples of the Talbot time $T_{\rm T}=2 M d^2/h$, where $M$ denotes the mass of the particles and $h$ is Planck's constant, the matter wave shows revivals.}
\label{fig:Talbot}
\end{figure}

\begin{figure}[t]
\begin{center}
\includegraphics[width=10cm,angle=0]{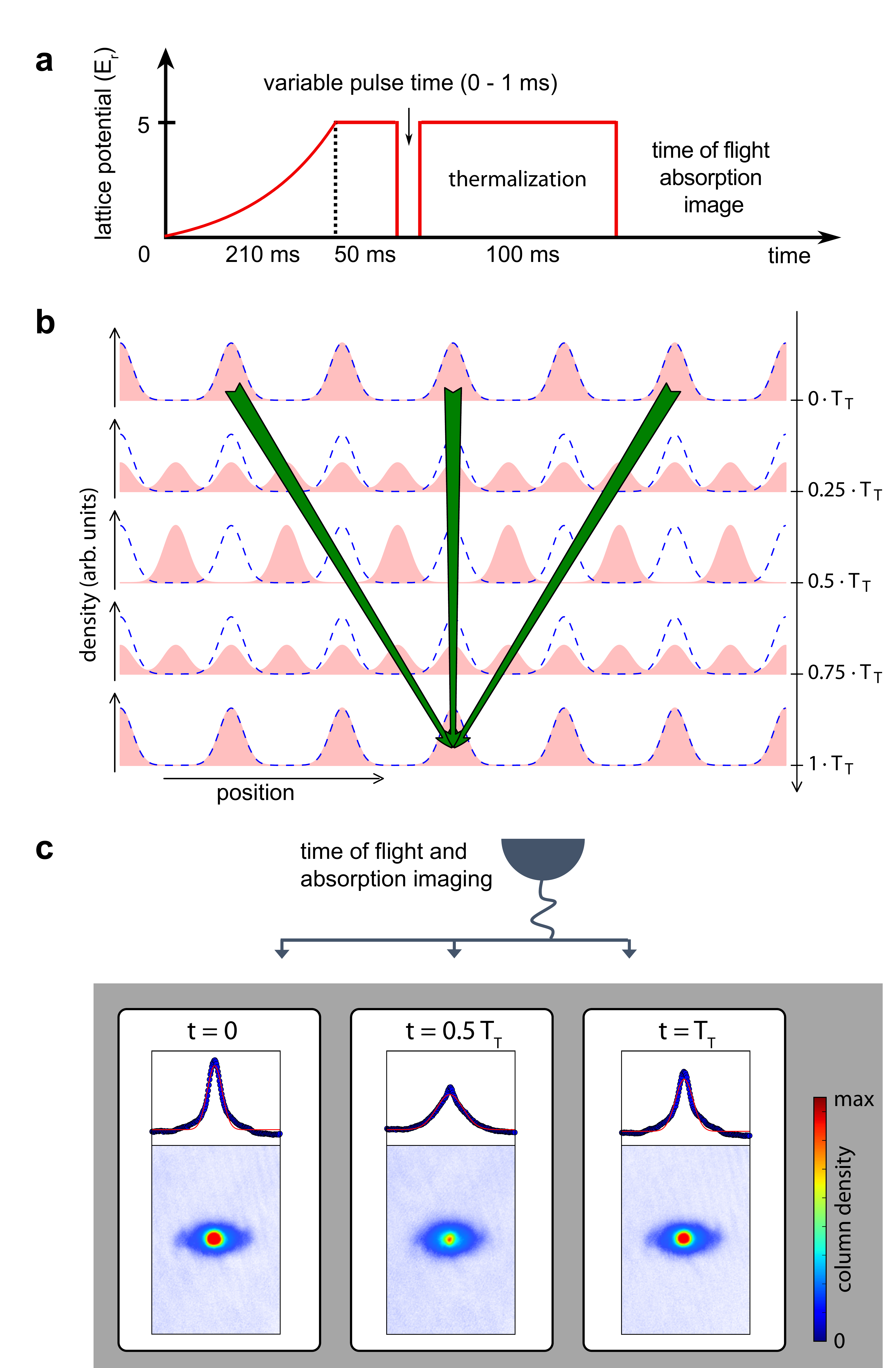}
\end{center}
\caption{{\bf Experimental protocol} {\bf a} Ultracold atoms are adiabatically loaded in an optical lattice potential. The red solid line indicates the height of the optical lattice potential, given in units of the recoil energy $E_{\rm r}$. After ramping up the lattice potential (vertical dashed black line), the atoms dwell for 50\,ms in the lattice potential. The potential is then blanked for a short time, during which the matter waves are allowed to freely expand along the lattice direction. Afterwards, the lattice potential is switched on again. After a subsequent hold time of 100\,ms, during which the atoms thermalize, the atomic density distribution is imaged in time of flight. {\bf b} Evolution of the density distribution of a coherent matter wave released from a periodic potential. After the Talbot time $T_{\rm T}=2 M d^2/h$, the initial density distribution is recovered. The arrows indicate the dominant contribution for the density in the central site after one Talbot time (see also text). In case of a partially coherent matter wave, the amplitude of the first revival will depend on the phase coherence between the next nearest neighbours. {\bf c} Time of flight absorption images and integrated line profiles for different evolution times. The red solid lines are Gaussian fits to the profiles.}
\label{fig:1}
\end{figure}

{\large {\bf Results}}

{\bf Near-field interferometry} The Talbot effect was first discovered in optics \cite{Talbot1836} and has later been applied to study matter wave interference with atomic and molecular beams \cite{Chapman1995,Brezger2002,Gerlich2007}. In a typical setup, the light or matter wave with wavelength $\lambda$ passes two consecutive gratings with lattice constant $d$, separated by the distance $L$. After integer multiples of the Talbot distance $L_{\rm T}=2d^2/\lambda$ one observes a self imaging of the wave and maximum transmission, see Fig.~\ref{fig:Talbot}a. Complementary, the Talbot effect can also be observed in the temporal domain studying ultracold atoms in optical lattices \cite{Deng1999,Mark2011}, see Fig.~\ref{fig:Talbot}b. The corresponding Talbot time is given by $T_{\rm T}=2Md^2/h$, where $M$ is the mass of the particles and $h$ is Planck's constant, and is connected to the Talbot distance via the deBroglie relation. 
Our near-field interferometer is based on the temporal Talbot effect and relies on a fast blanking of the lattice potential. Upon switching off, all lattice sites emit matter waves, which interfere with each other. After a variable time of free evolution, the lattice potential is switched on again and the matter wave is projected back onto the original lattice potential. The atoms are then allowed to thermalize. At integer multiples of the Talbot time, the atomic density distribution shows revivals, where the emerging contrast depends on the phase coherence between the interfering wave packets. Thereby, later revivals correspond to the interference of matter waves from more distant lattice sites.

We start describing the basic principle of the Talbot effect and its adaption to ultracold atoms in optical lattices. 
In the following, we treat the atoms in the tight-binding limit and neglect interactions. 
We will justify later on, why this is a good approximation in all practical cases.

The on-site wave function is approximated by a Gaussian with width $\sigma$
\begin{eqnarray}
\label{eq0}
\psi(x) = \frac{1}{\pi^{1/4} \sqrt{\sigma}} \exp\left(-\frac{x^2}{2\sigma^2} \right) \,.
\end{eqnarray}
For the sake of simplicity, we consider a one-dimensional array with lattice spacing $d$. The extension to higher dimensional cubic lattices is straight forward.
At each lattice site, which we enumerate with $n$, the matter wave can have an individual phase $\phi_n$, such that the total wave function reads
\begin{eqnarray}
\label{eq1}
\Psi(x,t=0) =\sum_{n=-\infty}^{\infty} \psi(x-nd) \exp\left(i\phi_n\right) \, .
\end{eqnarray}
Here, we have assumed $\sigma \ll d$, so that the wave packets do not overlap significantly and can be normalized individually. 
We are interested in the expectation values 
\begin{equation}
\label{c}
C_{n-n'}=\langle\exp\left[i(\phi_n-\phi_{n'}\right]\rangle \, ,
\end{equation}
which describe the phase correlation between atoms in the lattice sites $n$ and $n'$. 
Note that we further assume that the system is translationally invariant, such that the phase correlators $C_{n-n'}$ only depend on the relative distance $n-n'$ between the sites.
For identical phase factors, i.e.~$\phi_n=\phi$, we have $C_{n-n'}=1$ and equation (\ref{eq1}) describes a matter wave which is coherently spread over the entire lattice.
Values smaller than 1 correspond to partial coherence. By definition, we have $C_0=1$.
The interferometric sequence consists of a blanking of the optical lattice for a short time interval. 
Upon switching off the lattice potential, the matter wave starts to expand freely along the direction of the lattice. 
The momentum wave function is given by
\begin{widetext}
\begin{eqnarray}
\label{eq2}
\Phi(k,t) = \pi^{1/4} \sqrt{2 \sigma} \exp\left(-\frac{k^2\sigma^2}{2} - i\frac{\hbar k^2}{2 M}t\right) \sum_{n=-\infty}^{\infty} \exp\left( indk+i\phi_n\right)\,.
\end{eqnarray}
\end{widetext}
We first consider identical phase factors, i.e.~$\phi_n=\phi$, in all lattice sites, as it is the case in the ideal Talbot effect. The sum in equation (\ref{eq2}) can then be rewritten with the help of the Poisson sum formula 
(see equation (\ref{Poisson}) in the Methods) from Ref.~\cite{Kleinert2009} in terms of a Dirac comb, which has only components at integer multiples of the lattice vector $2\pi /d$. This yields
\begin{widetext}
\begin{eqnarray}
\label{eq2b}
\Phi(k,t) = \pi^{1/4} \sqrt{2 \sigma} \,\frac{2 \pi}{d} \,e^{i \phi } \sum_{m=-\infty}^{\infty} \delta \left( k- \frac{2 \pi}{d}m \right) 
\exp\left[- \left( \frac{2 \pi^2 \sigma^2}{d^2} + 2 \pi i \frac{t}{T_{\rm T}}\right) m^2 \right]\,,
\end{eqnarray}
\end{widetext}
where we have defined
the Talbot time $T_{\rm T}=2 M d^2/h$. We find that the wave function is restored at integer multiples $l$ of the Talbot time
\begin{equation}
\label{eq3}
\Phi(k,l T_{\rm T}) = \Phi(k,0).
\end{equation}
These periodic revivals of the matter wave field form the basis of the Talbot effect. Note that the presence of interactions, inhomogeneities, or fluctuations can damp or wash out these revivals.

In the above presented realization of the temporal Talbot effect, the time evolution is governed by the kinetic energy only, because no external force is present. The matter wave packets therefore interfere in real space but not in momentum space, where they simply acquire dynamical phase factors, which re-phase after the Talbot time. This is complementary to the realization of the temporal Talbot effect in Ref.\,\cite{Mark2011}, where the lattice potential is kept on, tunnelling is suppressed and an additional parabolic potential is applied. In this case, the matter wave packets interfere in momentum space, while they acquire in each lattice site phase factors, which re-phase after the Talbot time.

The interferometric sequence is completed by switching the optical lattice potential on again in a non-adiabatic way. 
Because the density distribution after the Talbot time $T_{\rm T}$ is identical to the initial one, it perfectly fits to the lattice potential.
Consequently, the lattice potential can be switched on without introducing additional potential energy to the atoms. 
For any other time, however, the density distribution is different from the initial one and the non-adiabatic loading in the lattice results in partially populating higher bands, thus adding potential and kinetic energy to the atoms.
For this reason, we introduce an additional hold time in our experimental sequence, during which the excess energy is converted into heating. 
Consequently, a measurement of either the band population immediately after the switching on of the lattice or of the temperature after relaxation is expected to show oscillations with revivals which are equal to integer multiples of the Talbot time. Fig.~2 summarizes the different steps of the interferometric sequence.

\begin{figure}[t]
\begin{center}
\includegraphics[width=8cm,angle=0]{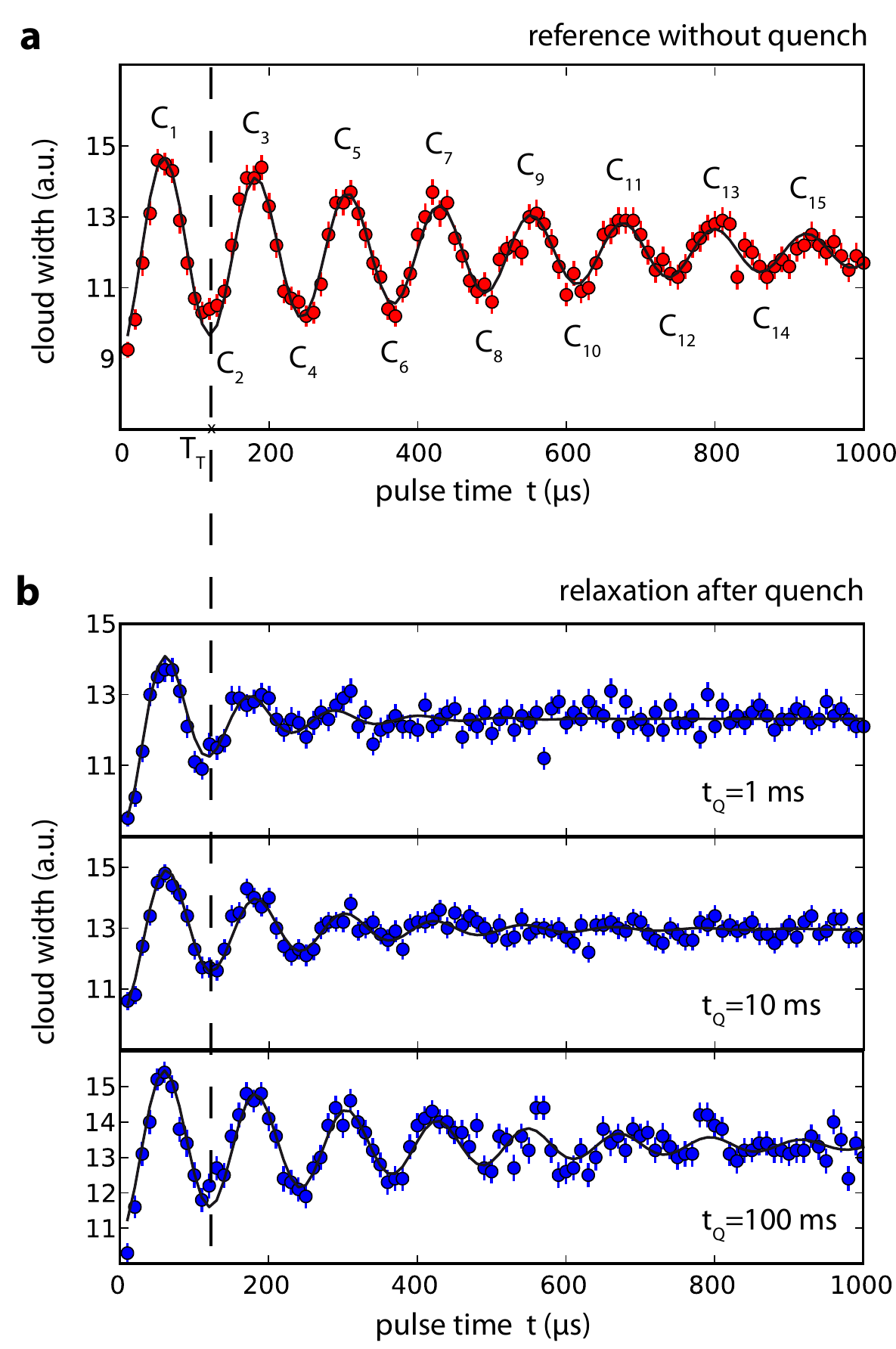}
\end{center}
\caption{{\bf Near-field interferometric measurement of  phase coherence in an optical lattice} {\bf a} A Bose-Einstein condensate is adiabatically loaded in a one-dimensional optical lattice with depth $s=5$. The peaks are labelled with the corresponding phase correlator $C_N$. The vertical dashed line denotes the Talbot time $T_{\rm T}=123\pm 1\,\mu$s. {\bf b} Same as (a) after performing a quantum quench from a deep lattice with $s=20$ to a lattice depth of $s=5$ within $500\mu\,$s. The interferometric sequence was started 1\,ms, 10\,ms, and
100\,ms after the quench. With increasing wait time, the interference pattern approaches that of the 
reference (a). The solid lines are fits with an exponentially damped sine. The error bars indicate the standard deviation of the mean of ten images per data point.}
\label{fig:2}
\end{figure}

An essential aspect of the Talbot effect is the near-field character of the interference pattern. 
This can be seen by Fourier transforming the momentum distribution in equation (\ref{eq2b}) back to position space:
\begin{widetext}
\begin{equation}
\label{eq4}
\Psi(x,\tau) =\frac{\sqrt{\sigma}}{\pi^{1/4} \sqrt{\sigma^2+id^2\tau/\pi}}  \sum_{n=-\infty}^{\infty}
\exp\left[-\frac{(x-nd)^2}{2(\sigma^2+id^2\tau/\pi)}  + i\phi_n  \right] \, .
\end{equation}
\end{widetext}
Here, $\tau=t/T_{\rm T}$ measures the time in units of the Talbot time. 
Upon switching on the lattice potential, the wave function (\ref{eq4}) is projected back onto the original array of Wannier functions (equation (\ref{eq0})). 
The resulting density overlap with one site is then given by
\begin{widetext}
\begin{eqnarray}
\label{eq5}
n_0(\tau)=|\braket{\psi(x)|\Psi(x,\tau)}|^2=\left| \sqrt{\frac{2\sigma^2}{2\sigma^2+id^2\tau/\pi}}
\sum_{n=-\infty}^{\infty} \exp\left[-\frac{d^2n^2}{2(2\sigma^2+id^2\tau/\pi)}+i\phi_n\right]\right|^2\, . 
\end{eqnarray}
\end{widetext}
Averaging over many experimental realizations, we find in the limit $d \gg \sigma$ (see Methods):
\begin{widetext}
\begin{eqnarray}
\label{dob}
\overline{n_0(\tau=N)} & \approx &   \frac{\sqrt{2 \pi} \sigma}{d} \sum_{n=-\infty}^{\infty} C_{2Nn} 
\exp \left( - \frac{2 \pi^2 \sigma^2 }{d^2 }\, n^2\right) \, , \\
\label{doc}
\overline{n_0(\tau=N+1/2)} &\approx &  \frac{\sqrt{2 \pi} \sigma}{d} \sum_{n=-\infty}^{\infty} (-1)^n C_{(2N+1)n} 
\exp \left( - \frac{ 2 \pi^2 \sigma^2}{d^2 } \, n^2\right) \, ,
\end{eqnarray}
\end{widetext}
where $N$ is an integer. At integer multiples of the Talbot time, equation (\ref{dob}) describes the revivals of the matter wave. 
In the presence of phase fluctuations, however, the phase correlators $C_{2Nn}$ are smaller than one, thus reducing the amplitude of the revivals. 
For half integer multiples of the Talbot time, equation (\ref{doc}) describes the density halfway between the revivals of the matter wave, which is illustrated in the
central picture in Fig.~\ref{fig:1}b.

For typical experimental parameters, the numerical factor in the exponent of equations (\ref{dob}) and (\ref{doc}) is close to 1. In our particular example, it amounts to $2\pi^2\sigma^2/d^2\approx 0.89$ (see Methods).
We can therefore restrict the sum to the first two non-trivial leading terms:
\begin{widetext}
\begin{eqnarray}
\label{dod}
\overline{n_0(\tau=N)} & \approx & \frac{\sqrt{2 \pi} \sigma}{d} \left[ 1 + 2 C_{2N}
\exp \left( - \frac{2 \pi^2 \sigma^2}{d^2} \right) + \ldots \right]\, ,\\
\label{doe}
\overline{n_0(\tau=N+1/2)} & \approx &  \frac{\sqrt{2 \pi} \sigma}{d} \left[ 1 - 2 C_{2N+1}
\exp \left( - \frac{2 \pi^2 \sigma^2}{d^2  } \right) + \ldots \right]\, .
\end{eqnarray}
\end{widetext}
The above expressions represent the central result of this work. 
They have a straightforward interpretation: the consecutive maxima and minima in the Talbot signal can be directly connected to the phase correlators between lattice sites with a well defined distance $(n-n')$. Therefore we can map the pulse time $t$ of the Talbot signal to a spatial coordinate for the phase correlation according to

\begin{equation}
\label{eq:mapping}
N= 2\times \frac{t}{T_{\rm T}},
\end{equation}

where $N=n-n'$ measures the distance in units of the lattice constant (see Fig.~3a). 
This fundamental relation enables the measurement of finite-range first-order correlations in an optical lattice.\\

{\bf Experiment} In the following, we apply this interferometry to experimentally measure the spreading of phase coherence in a one-dimensional optical lattice after a quantum quench. In a first experiment, we adiabatically load a cigar-shaped Bose-Einstein condensate of $^{87}$Rb 
atoms in a one-dimensional optical lattice of depth $s=5$. Each lattice site is occupied by a large number of atoms ($\approx 800$ in the trap center.) Details of the experimental setup can be found in the Methods. We then blank the lattice for time intervals between 0 and 1\,ms. 
Subsequently, we keep the atoms in the same lattice for another 100\,ms to allow for thermalization. 
We choose the total width of the density distribution of the time of flight absorption image (see Fig.\,2c) as the interferometer signal. 
The result is shown in Fig.~3a and reveals pronounced oscillations of the fitted cloud width.
We observe up to seven revivals of the matter wave field. Fitting an exponentially damped sine function, we find a Talbot time of $123\pm 1$\,\textmu s which is close to the theoretical value of 130\,\textmu s. The small deviation by about 5\,\%  might stem from a small misalignment in the experiment,
see also Methods. 
The contrast of the interferometer has a decay constant of $525\pm40\,$\textmu s. 
We attribute this decay to the presence of interactions in the system. For the chemical potential of $\mu = h \times 1.4\,$kHz, we can calculate an associated typical time scale of $h/\mu\approx 700\,\mu$s, on which interaction effects are expected to become important. A more detailed discussion of the influence of the interactions is given in the Method section.

It is remarkable that even in the presence of interactions, the Talbot effect can be observed.
This is because the Talbot effect acquires its kinetics from the localized Wannier function in the lattice, which expands on a time scale, which is about one order of magnitude
faster than the typical interaction-induced dynamics.
Therefore, treating the atoms non-interacting is indeed a good approximation for the first revivals.
This separation of time scales automatically emerges, whenever the quantum gas resides in the lowest band of the lattice potential.
The method is therefore applicable in many experiments with ultracold atoms in optical lattices.
  
As concluded from (\ref{dod}) and (\ref{doe}) 
each maximum and minimum in the Talbot signal corresponds to a consecutive phase correlator $C_N$.
In the present experiment, the maximum distance over which the phase correlations can be probed is about 14 lattice constants, see Fig.~3a.

So far we have not evaluated the absolute value of the Talbot signal in the minima and maxima. 
This is, indeed, a more complex task as it requires the precise modelling of how the lattice blanking converts into band occupation or deposited energy.
However, as we will show in the next paragraph, this is not a necessary requirement for measuring
finite-range phase coherence.\\

{\bf Spreading of phase correlations} We now study the quantum quench of a Bose-Einstein condensate residing in a one-dimensional optical lattice \cite{Cataliotti2001}. 
The quench is realized by adiabatically loading the BEC in a deep lattice with $s=20$, where tunnelling is strongly 
suppressed and only little phase coherence between the sites exists, and suddenly switching to a shallow lattice with $s=5$, where tunneling sets in and phase coherence starts spreading, see Fig.~4a for an illustration. 
The quench is done within 500\,\textmu s, which is much faster than the tunnelling time and slow enough to preserve the band occupancy. 
After a variable equilibration time $t_{\rm Q}$ in the shallow lattice, during which the phase coherence between the sites builds up, we perform the interferometric sequence as described above. 
Fig.~3b shows the measured interference pattern for three different equilibration times $t_{\rm Q}$. 
It is clearly visible how the later revivals gain more and more contrast for increasing $t_{\rm Q}$ and the reference without quantum quench of Fig.~3a is approached.
This gives a first qualitative picture of how the coherence spreads over the lattice.

\begin{figure}[t]
\begin{center}
\includegraphics[width=8cm]{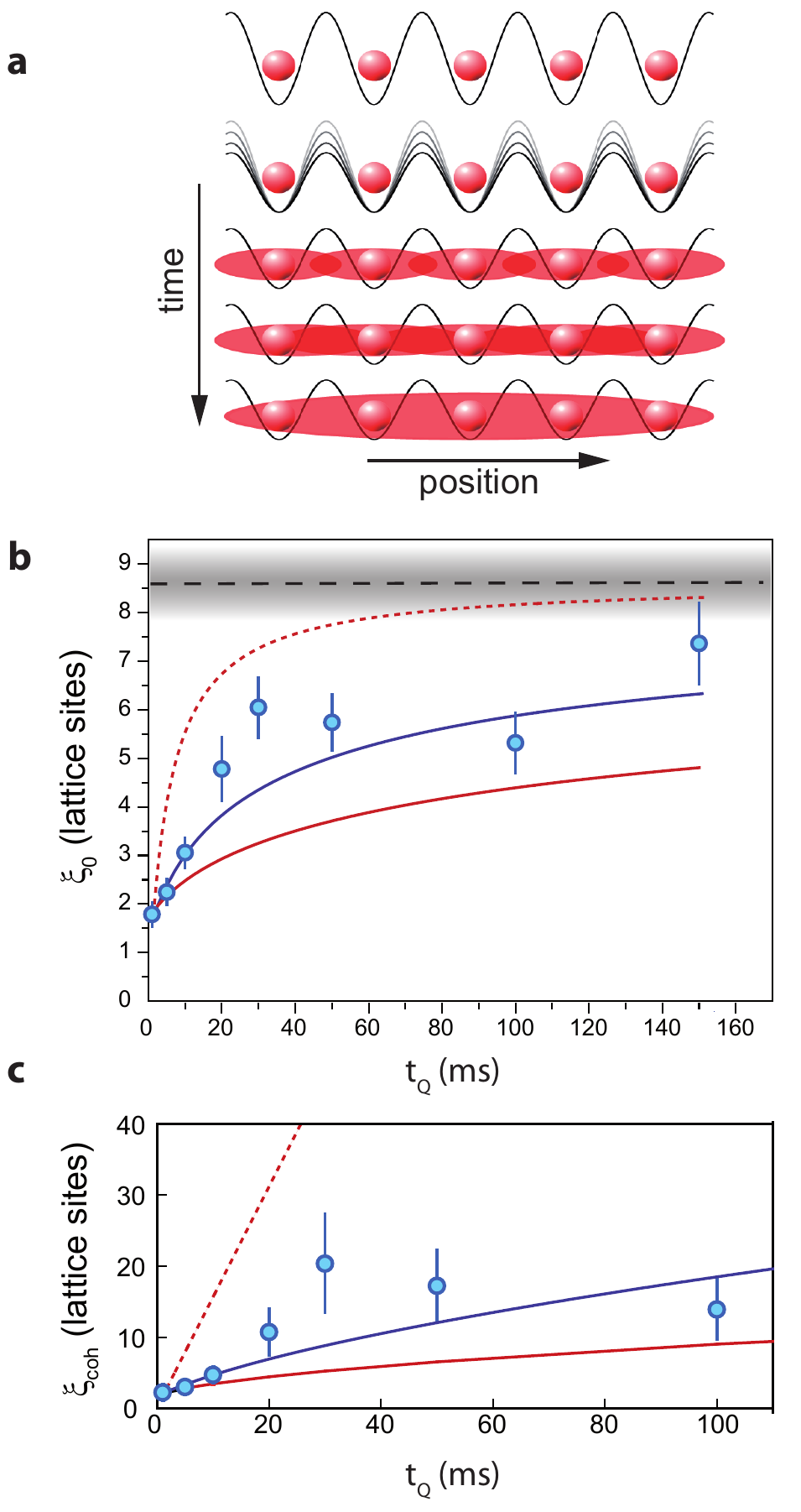}
\end{center}
\caption{{\bf Build-up of phase coherence} {\bf a} After lowering the lattice from $s=20$ to $s=5$, phase coherence between the lattice sites starts to develop. {\bf b} Temporal evolution of the spatial decay length $\xi_0$, derived from the Talbot signal shown in Fig.\,3b (blue dots). The black dashed line shows the decay of the reference in Fig.\,3a. The grey shaded area and the blue vertical bars denote the fit errors. {\bf c} Phase coherence length $\xi_{\rm coh}$, corrected for the decay of the reference (blue dots). The error bars are derived from {\bf b}. The blue solid line is a power law fit, yielding an exponent of $\alpha=0.6\pm 0.1$. The red dashed line indicates the fastest possible linear spread of phase coherence. The red solid line indicates the spread of phase coherence in a classical random walk model, scaling as $\propto \sqrt{t_{\rm Q}}$. All three curves are locked to the first data point. For comparison, the two models and the fit are also shown in (b), incorporating the decay of the reference.}
\label{fig:quench1D}
\end{figure}   

To further evaluate the data, we fit an exponentially damped sine to the individual Talbot signals. The results are shown in Fig.\,3b as solid lines, well matching the experimental data. We then convert the resulting decay time $t_{\rm T}$ into the spatial decay length $\xi_0=2t_{\rm T}/T_{\rm T}$ according to equation (\ref{eq:mapping}), where $\xi_0$ measures the distance in units of the lattice constant. The build-up of phase coherence after the quench can then be seen by an increase of the decay length with $t_{\rm Q}$, as shown in Fig.\,4b. Directly after the quench ($t_{\rm Q}=1\,$ms), we find residual correlations which extend over almost two lattice sites. We attribute this to an incomplete dephasing of the system during the preparation phase and a build-up of short range phase coherence during the first millisecond. The decay length increases up to $t_{\rm Q}=150\,$ms, where it overlaps within the error bar with that of the reference.

The reference system is well in the condensate phase, which exhibits long-range order and implies an infinite coherence length. As discussed above, the observed decay of the reference (Fig.\,3a) can be explained with the presence of interactions. We therefore assume in the following that the reference corresponds to a system with {\it infinite} coherence length. The measured exponential decay of the Talbot signal $\xi_0$ has now two contributions: the first one is the bare decay of the phase correlations, denoted by the coherence length $\xi_{\rm coh}$. The second one is the interaction induced decay, given by the decay length $\xi_{\rm ref}$ of the reference. Because both decay mechanisms act together, the corresponding constants add according to $1/ \xi_0=1/ \xi_{\rm coh}+1/ \xi_{\rm ref}$. Fig.\,4c shows the resulting coherence length $\xi_{\rm coh}$ in dependence of $t_{\rm Q}$. By fitting a power law (solid line), we find a scaling $\propto t_{\rm Q}^{\alpha}$ with the exponent $\alpha=0.6\pm 0.1$.

We now compare our results to theoretical models. In the system under investigation, i.e. a one-dimensional array of 2D condensates, the quantum quench induces elementary excitations and we expect the occurrence of phase slips or soliton-like excitations along the lattice. Due to the two transverse degrees of freedom in each lattice site, these excitations can further decay, for instance, in vortices. As a result, complex microscopic quantum dynamics set in. For long equilibration times $t_{\rm Q}$ however, we expect (and also observe) a system which is close to thermal equilibrium again and whose excitations have eventually relaxed.

A quantitative description of these complex non-equilibrium dynamics is not feasible due to the high occupation number in each lattice site. But two related, conceptually simpler scenarios were studied numerically in Ref.~\cite{Dziarmaga2012}: For a 3D Mott insulator which was quenched to a superfluid it was found that the coherence length shows a scaling $\propto \sqrt{t_{\rm Q}}$ after the quench. Furthermore, in a strictly one-dimensional scenario, a linear scaling $\propto t_{\rm Q}$  was detected. While our experimental results are closer to the prediction of the 3D quench scenario, none of the two scenarios is fully applicable here, as our system combines 1D, 2D, and 3D properties in a non-trivial way.

Easier to treat are the limiting cases for the spreading of phase correlations. The fastest possible speed at which excitations can move through the system is given by coherent tunnelling dynamics, where the coherence increases linearly in time. For a phase difference of $\pi/2$ between neighbouring sites, the group velocity of a matter wave becomes maximum, $v_{\rm max}=2dJ/\hbar$. Here, $J$ is the tunneling coupling and $\hbar/J\approx 1.3\,$ms is the tunnelling time for the given experimental parameters. This result applies for interacting and non-interacting systems and sets an upper velocity, with which the phase coherence can spread. A corresponding lower bound can be derived by considering a classical random walk. In this model, a particle transports the phase information and hops randomly through the lattice with the rate $J/\hbar$. The result is a diffusive motion, whose spatial width grows as $d\sqrt{J/\hbar}\sqrt{t_{\rm Q}}$. In Fig.\,4c, we show both limiting cases. The experimental data lie well within this corridor, being closer to the diffusive motion. This is also in accordance with the result for the power law exponent.

Our results hence indicate that the phase coherence after the quench does not build up light-cone like, as it was observed, e.g., for particle-hole excitations in a 1D Mott insulator background \cite{Cheneau2012}, but rather like a diffusive process. We know from own previous studies on the same experimental system \cite{Labouvie2015,Labouvie2016}, that the transverse degrees of freedom within each lattice site are important for the dynamics of our system (see also Methods). This renders the system dynamics three-dimensional, even though we eventually measure 1D phase coherence. This could explain, why we do not find a linear spread of phase coherence, as it is predicted in Ref.\,\cite{Dziarmaga2012}. For a more quantitative understanding of our results, however, a detailed microscopic modelling of the experimental situation would be mandatory.\\

{\large {\bf Discussion}} 

Several extensions of the near-field interferometer can be foreseen.
In higher-dimensional optical lattices, the phase coherence can be measured independently in all lattice directions by only switching the appropriate lattice axis.
Provided the read-out of the contrast observable can be made spatially resolved, such as in high resolution {\it in situ} detection experiments \cite{Gericke2008,Sherson2010,Bakr2010}, the technique can be used to measure locally the degree of phase coherence in an optical lattice. 
In combination with single site resolved quantum gases microscopy, even nearest neighbour phase correlations might become accessible.
How powerful such techniques can be, has been demonstrated in experiments with one-dimensional quantum gases \cite{Gring2012,Langen2015}.

The possibility to probe finite-range phase correlations in an ultracold lattice gas offers many perspectives for the study of ground-state properties and non-equilibrium dynamics, especially in the context of quantum quenches. 
The demonstrated near-field interferometer technique is general, versatile and can be applied in practically all lattice experiments without major modifications. 
It has the potential to become a standard diagnostic tool in experiments with optical lattices, complementing advanced techniques for the measurement of atomic densities.\\

{\large {\bf Methods}}



{\bf Experimental setup} In the experiment, we prepare a Bose-Einstein condensate of $^{87}$Rb atoms in a 1D optical lattice with a nominal lattice spacing of $d=547\,\rm{nm}$. 
The lattice is produced by two laser beams with wavelength $\lambda=774$\,nm and waist $w =500$\,\textmu m intersecting under an angle of $90\pm 2$ degrees.
The angle uncertainty, which is due to geometrical constraints of the setup, results in a systematic error of the Talbot time in the order of a few percent.
The transverse confinement is provided by an optical dipole trap of wavelength $\lambda=10.6$\,\textmu m with a trapping frequency of $\omega_{\perp}= 2\pi \times 170\,$s$^{-1}$. 
For a typical number of 50,000 atoms, about 800 atoms are residing in each lattice site.
The overall system can therefore be considered as a one-dimensional array of weakly coupled two-dimensional BECs \cite{Cataliotti2001}.
The chemical potential in the center of the trap is derived from the transverse extension of the atomic cloud in the central lattice sites and amounts to $\mu=h\times1.4\,$kHz.
The lattice depth $V_0$ is expressed in terms of the recoil energy $E_{\rm r} = \pi^2 \hbar^2 / (2 M d^2)$, where $M$ is the rubidium mass. For the interferometric sequence, we always chose $s=5$, for which we have $d\approx 5 \sigma$.
For this lattice depth, the oscillation period in the transverse direction is on the same order as the tunnelling coupling. Atoms tunneling from one site to another can partially explore the transverse degree of freedom before tunnelling into another site. 
The overall dynamics is therefore not one-dimensional but has rather 3D characteristics.  \\


{\bf Calculation of phase correlators} Here we show in detail that averaging the density overlap equation (\ref{eq5}) over many experimental
realizations yields in the limit $d \gg \sigma$ the equations (\ref{dob}) and (\ref{doc}).
To this end we
analyse the density overlap equation (\ref{eq5}) averaged over the phase fluctuations $\phi_n$, which yields 
\begin{eqnarray}
\overline{n_0(\tau)} = \frac{2 \sigma^2}{\sqrt{4\sigma^4+\frac{\tau^2 d^4}{\pi^2}}}
\sum_{n=-\infty}^{\infty}  \sum_{n'=-\infty}^{\infty}  C_{n-n'} \, 
\exp \left[ - \frac{\sigma^2 d^2}{4\sigma^4 + \tau^2 d^4/\pi^2} \left( n^2+ {n'}^2 \right) 
+\frac{i \tau d^4 / (2 \pi) }{4\sigma^4 + \tau^2 d^4/\pi^2}  \left( n^2 -  {n'}^2 \right) \right] \, .        
\label{ddsum}
\end{eqnarray}
Here we have introduced the phase correlator equation (\ref{c})
so that the phase fluctuations depend only on the distance between the lattice sites and are homogeneous over the extension of the cloud. 
Note that the resulting missing population of atoms in the initial ground state results in a population of higher bands and leads to heating during the hold time.

As a next step we rearrange the double sum in equation (\ref{ddsum})
 in the following fashion. We define new summation indices
$k=n+n'$ and $l=n-n'$, so that we have conversely $n=(k+l)/2$ and $n'=(k-l)/2$, which implies $n^2+ {n'}^2= (k^2 + l^2)/2$ and $n^2- {n'}^2=kl$.
Note that it is crucial to distinguish between two different cases with respect to the new summation indices. 
Either we could have both $k=2K$ and $l=2L$ to be even or $k=2K+1$ and $l=2L+1$ to be odd. As a result the density overlap in equation (\ref{ddsum}) decomposes into the two terms
\begin{eqnarray}
\label{do}
\overline{n_0(\tau)} = \sum_{L=-\infty}^{\infty} \Big[ C_{2L} F_{2L} (\tau) + C_{2L+1} F_{2L+1} (\tau) \Big] \, , 
\end{eqnarray}
where the respective weights of the phase correlators with an even number of lattice sites turns out to be
\begin{eqnarray}
\label{even}
F_{2L} (\tau) =\frac{2 \sigma^2}{\sqrt{4\sigma^4+\frac{\tau^2 d^4}{\pi^2}}} \exp \left[ - \frac{2 \sigma^2 d^2 L^2}{4\sigma^4 + \tau^2 d^4/\pi^2} \right] 
\sum_{K=-\infty}^{\infty}  \exp \left[ - \frac{2 \sigma^2 d^2 K^2}{4\sigma^4 + \tau^2 d^4/\pi^2} +\frac{2 i \tau d^4 LK/ \pi }{4\sigma^4 + \tau^2 d^4/\pi^2} \right] \, ,
\end{eqnarray}
whereas for an odd number of lattice sites we obtain
\begin{eqnarray}
F_{2L+1} (\tau)&=& \frac{2 \sigma^2}{\sqrt{4\sigma^4+\frac{\tau^2 d^4}{\pi^2}}} \exp \left[ - \frac{2 \sigma^2 d^2 \left( L +1/2 \right)^2}{4\sigma^4 + \tau^2 d^4/\pi^2} \right] 
\nonumber \\
&& \times \sum_{K=-\infty}^{\infty}  \exp \left[ - \frac{2 \sigma^2 d^2  \left( K +1/2 \right)^2}{4\sigma^4 + \tau^2 d^4/\pi^2} 
+\frac{2 i \tau d^4 \left( L +1/2 \right) \left( K +1/2 \right) / \pi }{4\sigma^4 + \tau^2 d^4/\pi^2} 
 \right] \, .
\label{odd}
\end{eqnarray}
As a cross-check we observe that, indeed, $\overline{n_0(0)}=1$ due to $d \gg \sigma$. The series for the respective weights of the phase correlators in equations (\ref{even}) and (\ref{odd}) can also be dual transformed.
To this end we use the Poisson sum formula \cite{Kleinert2009}, 
\begin{eqnarray}
\label{Poisson}
\sum_{m=-\infty}^{\infty} \delta(x-m) = \sum_{n=-\infty}^{\infty} \exp\left( 2 \pi i n x \right) \, .
\end{eqnarray}
which implies the dual transformation of a series
\begin{eqnarray}
\label{DT}
\sum_{m=-\infty}^{\infty} f(m) = \sum_{n=-\infty}^{\infty} \int_{- \infty}^{+ \infty} d x \, f(x) \, \exp\left( 2 \pi i n x \right) \, .
\end{eqnarray}
Applying equation (\ref{DT}) yields for the weights of the phase correlators in equation (\ref{even})
\begin{eqnarray}
\label{evenb}
F_{2L} (\tau) = \frac{\sqrt{2 \pi} \sigma}{d} \exp \left[ 
- \frac{(2 L)^2}{\frac{2 \tau^2 d^2}{\pi^2 \sigma^2} \left( 1 + \frac{4 \pi^2 \sigma^4}{\tau^2 d^4}\right) }
\right]\sum_{n=-\infty}^{\infty} \exp \left\{
- \frac{\left[ L - \left(1+ \frac{4 \pi^2 \sigma^4}{\tau^2 d^4}\right) \tau n  \right]^2}{\frac{2 \sigma^2}{d^2} \left( 1 + \frac{4 \pi^2 \sigma^4}{\tau^2 d^4}\right)}
\right\}
\end{eqnarray}
and, correspondingly, for (\ref{odd})
\begin{eqnarray}
\label{oddb}
F_{2L+1} (\tau) = \frac{\sqrt{2 \pi} \sigma}{d} \exp \left[ 
- \frac{(2 L+1)^2}{\frac{2 \tau^2 d^2}{\pi^2 \sigma^2} \left( 1 + \frac{4 \pi^2 \sigma^4}{\tau^2 d^4}\right) }
\right]\sum_{n=-\infty}^{\infty} (-1)^n \exp \left\{
- \frac{\left[ L +\frac{1}{2}- \left(1+ \frac{4 \pi^2 \sigma^4}{\tau^2 d^4}\right) \tau n  \right]^2}{\frac{2 \sigma^2}{d^2} \left( 1 + \frac{4 \pi^2 \sigma^4}{\tau^2 d^4}\right)}
\right\} \, .
\end{eqnarray}

In order to quantify this idea further,
we observe that,
when the propagation time coincides with integer and half-integer multiples of the Talbot time, the density overlap (equation (\ref{do})) 
with the weights equations (\ref{evenb}) and (\ref{oddb})
reduces to equations (\ref{dob}) and (\ref{doc}) due to the limit $d \gg \sigma$.\\

{\bf Interaction induced decay of the Talbot signal} We here present a semi-quantitative derivation, why the interaction between the atoms, characterized by the chemical potential $\mu$, induces a decay of the Talbot signal during the interferometric sequence. We start with considering the normalized axial density $n(x)$ in a single lattice site, given by the square of the Wannier function equation (\ref{eq0}). If we ignore an interaction induced broadening of the on-site wave function, the interaction energy in a mean field description is given by $E_{\rm int}(x)=\mu n(x)$. Upon switching off the lattice potential, the atoms experience a force $F(x)=-E'_{\rm int}(x)$, generated by the gradient of the interaction energy, which acts symmetrically on both sites. We estimate an average force, which pushes the atoms to one side by 

\begin{equation}
\overline{F}=\frac{\int_0^\infty F(x) n(x) dx}{\int_0^\infty n(x) dx} = \frac{\mu}{\pi \sigma}.
\end{equation}

After switching off the optical lattice, the momentum wave function consists initially of a series of $\delta$-functions (see equation (\ref{eq2b}) evaluated at $t=0$), which are separated by the width of the Brillouin zone $\Delta p=2\hbar \pi/d$. The interaction induced force leads to a symmetric broadening of each momentum peak. We now assume that half of this force, i.e. $\overline{F}/2$, is constantly present during the Talbot sequence. This can be justified by an inspection of the density distribution in Fig.\,2b at different times. While at $t=0.5\,T_{\rm T}$ and $t=T_{\rm T}$, the average force is the same as calculated above, it is only half as strong at times $t=0.25\,T_{\rm T}$ and $t=0.75\,T_{\rm T}$. For times in between, the density distribution is even smoother. After the time $t=\Delta p /\overline{F}$, each momentum peak has broadened over the whole Brillouin zone. Consequently, the momentum peaks start to overlap and the interference gets lost. Thus, the observation of the temporal Talbot effect is restricted due to interaction effects to a time interval of about $t=450\,\mu$s. This is compatible with the measured decay time of the reference of $525\,\mu$s.

Regarding the harmonic confinement along the lattice direction, its influence on the Talbot effect should be negligible. The largest distance over which we can measure phase correlations amounts to 14 lattice sites. The sample itself extends over 150 lattice sites. Thus, atoms interfering in the trap center over a distance of 14 lattice sites explore a potential energy shift of less than 10 percent of the chemical potential. The associated time scale is on the order of 10 - 15 ms, more than one order of magnitude slower than the observed decay time of the reference.\\

{\bf Data availability} The data that support the findings of this study are available
from the corresponding author upon request.


\begin{thebibliography}{10}
\expandafter\ifx\csname url\endcsname\relax
  \def\url#1{\texttt{#1}}\fi
\expandafter\ifx\csname urlprefix\endcsname\relax\def\urlprefix{URL }\fi
\providecommand{\bibinfo}[2]{#2}
\providecommand{\eprint}[2][]{\url{#2}}

\bibitem{Andrews1997}
\bibinfo{author}{Andrews, M.~R.} \emph{et~al.}
\newblock \bibinfo{title}{Observation of interference between two {B}ose
  condensates}.
\newblock \emph{\bibinfo{journal}{Science}} \textbf{\bibinfo{volume}{275}},
  \bibinfo{pages}{637-641} (\bibinfo{year}{1997}).

\bibitem{Greiner2001}
\bibinfo{author}{Greiner, M.}, \bibinfo{author}{Bloch, I.},
  \bibinfo{author}{Mandel, O.}, \bibinfo{author}{H\"ansch, T.~W.} \&
  \bibinfo{author}{Esslinger, T.}
\newblock \bibinfo{title}{Exploring phase coherence in a 2{D} lattice of
  {B}ose-{E}instein condensates}.
\newblock \emph{\bibinfo{journal}{Phys. Rev. Lett.}}
  \textbf{\bibinfo{volume}{87}}, \bibinfo{pages}{160405}
  (\bibinfo{year}{2001}).

\bibitem{Gerbier2005}
\bibinfo{author}{Gerbier, F.} \emph{et~al.}
\newblock \bibinfo{title}{Phase coherence of an atomic {M}ott insulator}.
\newblock \emph{\bibinfo{journal}{Phys. Rev. Lett.}}
  \textbf{\bibinfo{volume}{95}}, \bibinfo{pages}{050404}
  (\bibinfo{year}{2005}).

\bibitem{Braun2015}
\bibinfo{author}{Braun, S.} \emph{et~al.}
\newblock \bibinfo{title}{Emergence of coherence and the dynamics of quantum
  phase transitions}.
\newblock \emph{\bibinfo{journal}{Proc. Nat. Ac. Sci.}}
  \textbf{\bibinfo{volume}{112}}, \bibinfo{pages}{3641-3646} (\bibinfo{year}{2015}).

\bibitem{Ozeri2005}
\bibinfo{author}{Ozeri, R.}, \bibinfo{author}{Katz, N.},
  \bibinfo{author}{Steinhauer, J.} \& \bibinfo{author}{Davidson, N.}
\newblock \bibinfo{title}{\textit{Colloquium}: Bulk {B}ogoliubov excitations
  in a {B}ose-{E}instein condensate}.
\newblock \emph{\bibinfo{journal}{Rev. Mod. Phys.}}
  \textbf{\bibinfo{volume}{77}}, \bibinfo{pages}{187-205} (\bibinfo{year}{2005}).

\bibitem{Foelling2005}
\bibinfo{author}{F\"olling, S.} \emph{et~al.}
\newblock \bibinfo{title}{Spatial quantum noise interferometry in expanding
  ultracold atom clouds}.
\newblock \emph{\bibinfo{journal}{Nature}} \textbf{\bibinfo{volume}{434}},
  \bibinfo{pages}{481-484} (\bibinfo{year}{2005}).

\bibitem{Vassen2012}
\bibinfo{author}{Vassen, W.} \emph{et~al.}
\newblock \bibinfo{title}{Cold and trapped metastable noble gases}.
\newblock \emph{\bibinfo{journal}{Rev. Mod. Phys.}}
  \textbf{\bibinfo{volume}{84}}, \bibinfo{pages}{175-210} (\bibinfo{year}{2012}).

\bibitem{Jeltes2007}
\bibinfo{author}{Jeltes, T.} \emph{et~al.}
\newblock \bibinfo{title}{Comparison of the {H}anbury {B}rown-{T}wiss effect
  for bosons and fermions}.
\newblock \emph{\bibinfo{journal}{Nature}} \textbf{\bibinfo{volume}{445}},
  \bibinfo{pages}{402-405} (\bibinfo{year}{2007}).

\bibitem{Ott2016}
\bibinfo{author}{Ott, H.}
\newblock \bibinfo{title}{Single atom detection in ultacold quantum gases: a
  review of current progress}.
\newblock \emph{\bibinfo{journal}{Rep. Prog. Phys.}}
  \textbf{\bibinfo{volume}{79}}, \bibinfo{pages}{054401}
  (\bibinfo{year}{2016}).

\bibitem{Guarrera_2012}
\bibinfo{author}{Guarrera, V.} \emph{et~al.}
\newblock \bibinfo{title}{Spatiotemporal fermionization of strongly interacting
  one-dimensional bosons}.
\newblock \emph{\bibinfo{journal}{Phys. Rev. A}} \textbf{\bibinfo{volume}{86}},
  \bibinfo{pages}{021601} (\bibinfo{year}{2012}).

\bibitem{Cheneau2012}
\bibinfo{author}{Cheneau, M.} \emph{et~al.}
\newblock \bibinfo{title}{Light-cone-like spreading of correlations in a
  quantum many-body system}.
\newblock \emph{\bibinfo{journal}{Nature}} \textbf{\bibinfo{volume}{481}},
  \bibinfo{pages}{484-487} (\bibinfo{year}{2012}).

\bibitem{Hofferberth2007}
\bibinfo{author}{Hofferberth, S.}, \bibinfo{author}{Lesanovsky, I.},
  \bibinfo{author}{Fischer, B.}, \bibinfo{author}{Schumm, T.} \&
  \bibinfo{author}{Schmiedmayer, J.}
\newblock \bibinfo{title}{Non-equilibrium coherence dynamics in one-dimensional
  {B}ose gases}.
\newblock \emph{\bibinfo{journal}{Nature}} \textbf{\bibinfo{volume}{449}},
  \bibinfo{pages}{324-327} (\bibinfo{year}{2007}).

\bibitem{Hofferberth2008}
\bibinfo{author}{Hofferberth, S.} \emph{et~al.}
\newblock \bibinfo{title}{Probing quantum and thermal noise in an interacting
  many-body system}.
\newblock \emph{\bibinfo{journal}{Nat. Phys.}} \textbf{\bibinfo{volume}{4}},
  \bibinfo{pages}{489-495} (\bibinfo{year}{2008}).

\bibitem{Gring2012}
\bibinfo{author}{Gring, M.} \emph{et~al.}
\newblock \bibinfo{title}{Relaxation and prethermalization in an isolated
  quantum system}.
\newblock \emph{\bibinfo{journal}{Science}} \textbf{\bibinfo{volume}{337}},
  \bibinfo{pages}{1318-1322} (\bibinfo{year}{2012}).

\bibitem{Langen2015}
\bibinfo{author}{Langen, T.} \emph{et~al.}
\newblock \bibinfo{title}{Experimental observation of a generalized {G}ibbs
  ensemble}.
\newblock \emph{\bibinfo{journal}{Science}} \textbf{\bibinfo{volume}{348}},
  \bibinfo{pages}{207-211} (\bibinfo{year}{2015}).

\bibitem{Chomaz2015}
\bibinfo{author}{Chomaz, L.} \emph{et~al.}
\newblock \bibinfo{title}{Emergence of coherence in a uniform
  quasi-two-dimensional {B}ose gas}.
\newblock \emph{\bibinfo{journal}{Nat. Commun.}} \textbf{\bibinfo{volume}{6}},
  \bibinfo{pages}{6162} (\bibinfo{year}{2015}).

\bibitem{Bakr2010}
\bibinfo{author}{Bakr, W.~S.} \emph{et~al.}
\newblock \bibinfo{title}{Probing the superfluid-to-{M}ott insulator transition
  at the single-atom level}.
\newblock \emph{\bibinfo{journal}{Science}} \textbf{\bibinfo{volume}{329}},
  \bibinfo{pages}{547-550} (\bibinfo{year}{2010}).

\bibitem{Sherson2010}
\bibinfo{author}{Sherson, J.~F.} \emph{et~al.}
\newblock \bibinfo{title}{Single-atom-resolved fluorescence imaging of an
  atomic {M}ott insulator}.
\newblock \emph{\bibinfo{journal}{Nature}} \textbf{\bibinfo{volume}{467}},
  \bibinfo{pages}{68-72} (\bibinfo{year}{2010}).

\bibitem{Cataliotti2001}
\bibinfo{author}{Cataliotti, F.~S.} \emph{et~al.}
\newblock \bibinfo{title}{Josephson junction arrays with {B}ose-{E}instein
  condensates}.
\newblock \emph{\bibinfo{journal}{Science}} \textbf{\bibinfo{volume}{293}},
  \bibinfo{pages}{843-846} (\bibinfo{year}{2001}).

\bibitem{Talbot1836}
\bibinfo{author}{Talbot, H.~F.}
\newblock \bibinfo{title}{Facts relating to optical science. No IV}.
\newblock \emph{\bibinfo{journal}{Philos. Mag.}} \textbf{\bibinfo{volume}{9}},
  \bibinfo{pages}{401-407} (\bibinfo{year}{1836}).

\bibitem{Chapman1995}
\bibinfo{author}{Chapman, M.~S.} \emph{et~al.}
\newblock \bibinfo{title}{Near-field imaging of atom diffraction gratings: The
  atomic {T}albot effect}.
\newblock \emph{\bibinfo{journal}{Phys. Rev. A}} \textbf{\bibinfo{volume}{51}},
  \bibinfo{pages}{14(R)} (\bibinfo{year}{1995}).

\bibitem{Brezger2002}
\bibinfo{author}{Brezger, B.} \emph{et~al.}
\newblock \bibinfo{title}{Matter-wave interferometer for large molecules}.
\newblock \emph{\bibinfo{journal}{Phys. Rev. Lett.}}
  \textbf{\bibinfo{volume}{88}}, \bibinfo{pages}{100404}
  (\bibinfo{year}{2002}).

\bibitem{Gerlich2007}
\bibinfo{author}{Gerlich, S.} \emph{et~al.}
\newblock \bibinfo{title}{A {K}apitza-{D}irac-{T}albot-{L}au interferometer for
  highly polarizable molecules}.
\newblock \emph{\bibinfo{journal}{Nat. Phys.}} \textbf{\bibinfo{volume}{3}},
  \bibinfo{pages}{711-715} (\bibinfo{year}{2007}).

\bibitem{Deng1999}
\bibinfo{author}{Deng, L.} \emph{et~al.}
\newblock \bibinfo{title}{Temporal, matter-wave-dispersion {T}albot effect}.
\newblock \emph{\bibinfo{journal}{Phys. Rev. Lett.}}
  \textbf{\bibinfo{volume}{83}}, \bibinfo{pages}{5407} (\bibinfo{year}{1999}).

\bibitem{Mark2011}
\bibinfo{author}{Mark, M.~J.} \emph{et~al.}
\newblock \bibinfo{title}{Demonstration of the temporal matter-wave {T}albot
  effect for trapped matter waves}.
\newblock \emph{\bibinfo{journal}{New J. Phys.}} \textbf{\bibinfo{volume}{13}},
  \bibinfo{pages}{085008} (\bibinfo{year}{2011}).

\bibitem{Kleinert2009}
\bibinfo{author}{Kleinert, H.}
\newblock \emph{\bibinfo{title}{Path Integrals in Quantum Mechanics,
  Statistics, Polymer Physics, and Financial Markets, 5th Edition}}
  (\bibinfo{publisher}{World Scientific, Singapore}, \bibinfo{year}{2009}).

\bibitem{Dziarmaga2012}
\bibinfo{author}{Dziarmaga, J.}, \bibinfo{author}{Tylutki, M.} \&
  \bibinfo{author}{Zurek, W.~H.}
\newblock \bibinfo{title}{Quench from {M}ott insulator to superfluid}.
\newblock \emph{\bibinfo{journal}{Phys. Rev. B}} \textbf{\bibinfo{volume}{86}},
  \bibinfo{pages}{144521} (\bibinfo{year}{2012}).

\bibitem{Labouvie2015}
\bibinfo{author}{Labouvie, R.}, \bibinfo{author}{Santra, B.},
  \bibinfo{author}{Heun, S.}, \bibinfo{author}{Wimberger, S.} \&
  \bibinfo{author}{Ott, H.}
\newblock \bibinfo{title}{Negative differential conductivity in an interacting
  quantum gas}.
\newblock \emph{\bibinfo{journal}{Phys. Rev. Lett.}}
  \textbf{\bibinfo{volume}{115}}, \bibinfo{pages}{050601}
  (\bibinfo{year}{2015}).

\bibitem{Labouvie2016}
\bibinfo{author}{Labouvie, R.}, \bibinfo{author}{Santra, B.},
  \bibinfo{author}{Heun, S.} \& \bibinfo{author}{Ott, H.}
\newblock \bibinfo{title}{Bistability in a driven-dissipative superfluid}.
\newblock \emph{\bibinfo{journal}{Phys. Rev. Lett.}}
  \textbf{\bibinfo{volume}{116}}, \bibinfo{pages}{235302}
  (\bibinfo{year}{2016}).

\bibitem{Gericke2008}
\bibinfo{author}{Gericke, T.}, \bibinfo{author}{W{\"u}rtz, P.},
  \bibinfo{author}{Reitz, D.}, \bibinfo{author}{Langen, T.} \&
  \bibinfo{author}{Ott, H.}
\newblock \bibinfo{title}{High-resolution scanning electron microscopy of an
  ultracold quantum gas}.
\newblock \emph{\bibinfo{journal}{Nat. Phys.}} \textbf{\bibinfo{volume}{4}},
  \bibinfo{pages}{949-953} (\bibinfo{year}{2008}).\\\\

\end{thebibliography}


{\em Acknowledgments.} We would like to thank T. Langen and A. M{\"u}llers for valuable discussions. 
We acknowledge financial support by the German Research Foundation (DFG) within the SFB/TR 49 and the SFB/TR 185. 
B.S. is funded through a research grant of the University of Kaiserslautern,
R.L. and C.B. are supported by the graduate school of excellence MAINZ. A.B.B. acknowledges support from the German Academic Exchange Service (DAAD) and the University Grants Commission-Faculty Recharge Programme.\\

{\em Author contributions.} B.S. and H.O. conceived the experiment. H.O. supervised the experiment. H.O. and A.P. developed the theoretical model. B.S., C.B. and R.L. performed the experiment. B.S. and H.O. analyzed the data. A.B.B., B.S., C.B., H.O., R.L. and A.P. discussed the results and prepared the manuscript.\\

{\em Competing financial interests.} The authors declare no competing financial interests.


\end{document}